\documentclass[]{elsarticle} 
\usepackage[hyphens]{url}

  \journal{Epidemics 7} 

\usepackage{lineno} 
\providecommand{\tightlist}{%
  \setlength{\itemsep}{0pt}\setlength{\parskip}{0pt}}

\biboptions{sort&compress} 
\usepackage{graphicx}
\usepackage{booktabs} 

\usepackage[T1]{fontenc}
\usepackage{lmodern}
\usepackage{amssymb,amsmath}
\usepackage{ifxetex,ifluatex}
\usepackage{fixltx2e} 
\IfFileExists{upquote.sty}{\usepackage{upquote}}{}
\ifnum 0\ifxetex 1\fi\ifluatex 1\fi=0 
  \usepackage[utf8]{inputenc}
\else 
  \usepackage{fontspec}
  \ifxetex
    \usepackage{xltxtra,xunicode}
  \fi
  \defaultfontfeatures{Mapping=tex-text,Scale=MatchLowercase}
  
\fi
\IfFileExists{microtype.sty}{\usepackage{microtype}}{}
\ifxetex
  \usepackage[setpagesize=false, 
              unicode=false, 
              xetex]{hyperref}
\else
  \usepackage[unicode=true]{hyperref}
\fi
\hypersetup{breaklinks=true,
            bookmarks=true,
            pdfauthor={},
            pdftitle={Inferring HIV incidence trends and transmission dynamics with a spatio-temporal HIV epidemic model},
            colorlinks=true,
            urlcolor=blue,
            linkcolor=magenta,
            pdfborder={0 0 0}}
\usepackage{setspace}

\graphicspath{ {.} }

\begin{document}
\begin{frontmatter}

  \title{Inferring HIV incidence trends and transmission dynamics with a
spatio-temporal HIV epidemic model}
    \author[Imperial Maths]{Timothy M Wolock\corref{c1}}
   \ead{t.wolock18@imperial.ac.uk} 
   \cortext[c1]{Corresponding Author}
    \author[Imperial Maths]{Seth R Flaxman}
   \ead{s.flaxman@imperial.ac.uk} 
  
    \author[Imperial DIDE]{Jeffrey W Eaton}
   \ead{jeffrey.eaton@imperial.ac.uk} 
  
      \address[Imperial Maths]{Department of Mathematics, Imperial College London, London, UK}
    \address[Imperial DIDE]{Department of Infectious Disease Epidemiology, Imperial College London,
London, UK}
  
  \begin{abstract}
  Reliable estimation of spatio-temporal trends in population-level HIV
  incidence is becoming an increasingly critical component of HIV
  prevention policy-making. However, direct measurement is nearly
  impossible. Current, widely used models infer incidence from survey and
  surveillance seroprevalence data, but they require unrealistic
  assumptions about spatial independence across spatial units.
  
  In this study, we present an epidemic model of HIV that explicitly
  simulates the spatial dynamics of HIV over many small, interacting areal
  units. By integrating all available population-level data, we are able
  to infer not only spatio-temporally varying incidence, but also ART
  initiation rates and patient counts. Our study illustrates the
  feasibility of applying compartmental models to larger inferential
  problems than those to which they are typically applied, as well as the
  value of ``data fusion'' approaches to infectious disease modeling.
  \end{abstract}
   \begin{keyword} disease modeling, Bayesian inference, HIV modeling\end{keyword}
 \end{frontmatter}

\begin{center}\rule{0.5\linewidth}{\linethickness}\end{center}

\hypertarget{introduction}{%
\section{Introduction}\label{introduction}}

Fitting population-level models of the human immunodeficiency virus
(HIV) epidemic to a combination of survey and surveillance data is an
essential component of HIV policymaking (Stover et al.
\protect\hyperlink{ref-stover_updates_2017}{2017}). HIV is most
prevalent in countries with limited capacity for health surveillance, so
data are relatively sparse in the places where we need it most. We use
models to fill in the gaps and estimate the indicators we need in order
to assess the epidemic's trajectory (Brown et al.
\protect\hyperlink{ref-brown_improvements_2014}{2014}). These indicators
include things like adult HIV prevalence, HIV incidence, and coverage of
life-saving antiretroviral treatment (ART) (``Monitoring, Evaluation,
and Reporting (MER 2.0) Indicator Reference Guide''
\protect\hyperlink{ref-noauthor_monitoring_2017}{2017}).

Recently, improved data collection and expanded computational power have
increased demand for indicators at subnational levels (Meyer-Rath et al.
\protect\hyperlink{ref-meyer-rath_targeting_2018}{2018}). We have the
data and models to produce reliable maps of HIV prevalence, but similar
deep thought has not been given to inferring HIV incidence over space
(Dwyer-Lindgren et al.
\protect\hyperlink{ref-dwyer-lindgren_mapping_2019}{2019}; Cuadros et
al. \protect\hyperlink{ref-cuadros_mapping_2017}{2017}; Gutreuter et al.
\protect\hyperlink{ref-gutreuter_improving_2019}{2019}).Existing methods
for inferring HIV inference invariably treat each spatial unit as
separate from all others (Brown et al.
\protect\hyperlink{ref-brown_improvements_2014}{2014}; Stover et al.
\protect\hyperlink{ref-stover_updates_2017}{2017}), an obviously
incorrect assumption. This is more than a theoretical concern. If we
were to enact a truly effective HIV prevention programme in an urban
area and evaluate it without taking into account the new infections
prevented in surrounding suburban areas, we could easily underestimate
our programme's efficacy.

Our goal is to develop a model of HIV that can infer incidence over
space, as well as time. It must offer estimates at useful geographic
resolutions without incurring extreme computational costs. In this
technical report, we will provide a detailed account of our progress
towards this end.

First, we will outline the available population-level data sources that
might inform estimates of HIV incidence and distributional assumptions
we could use to integrate these data into a statistical model. Then, we
will describe the mechanistic model we are using to project the HIV
epidemic given a set of inferred parameters. Finally, we will provide
preliminary estimates from an application to Malawi and a summary of our
findings thus far.

\hypertarget{methodology}{%
\section{Methodology}\label{methodology}}

Because a new HIV infection might not be detected for several years,
incidence is virtually impossible to observe directly. Instead, we can
use mechanistic models to ``fuse'' together what indicators we do have
and infer the incidence series that was most likely to have generated
the combined observed dataset. This is the principle of data fusion
(Hall and Llinas \protect\hyperlink{ref-hall_introduction_1997}{1997}).
In this section, we will first describe and develop notation for types
of data used in our analysis and propose distributional assumptions that
we believe represent the generation processes of those data. Then we
will describe our compartmental model of HIV in detail.

\hypertarget{data-and-distributional-assumptions}{%
\subsection{Data and Distributional
Assumptions}\label{data-and-distributional-assumptions}}

We have identified three measurable indicators that will allow us to
infer HIV incidence: prevalence, ART coverage, and the proportion of
infections at a given time classified as ``recent.''

\hypertarget{prevalence}{%
\subsubsection{Prevalence}\label{prevalence}}

HIV antibody assay results are far and away the most prevalent and
reliable source of population-level HIV data. They come from two main
sources: large, nationally representative household surveys and routine
surveillance at antenatal care (ANC) clinics. These assays give us
seroprevalence data consisting the number of tests conducted in a given
group and the number of those tests that were positive. Let \(r\) and
\(t\) represent a geographic region and point in time, respectively.
Then for a given data source, \(s\), we can denote the number of tests
in region \(r\) at time \(t\) as \(T_{r,t}^{s, \text{HIV}}\) and the
number of positive tests as \(P_{r,t}^{s,\text{HIV}}\). Here, \(s\)
represents either a specific survey or a specific antenatal care site.

We assume that large household surveys are representative for every
region, so if \(s\) is a household survey,
\(P_{r,t}^{s,\text{HIV}}/T_{r,t}^{s,\text{HIV}}\) provides an unbiased
estimate of true prevalence in demographic segment \(\{r,t\}\), denoted
\(\rho_{r}(t)\). Therefore, we can assume that
\(P_{r,t}^{s,\text{HIV}}\) is a sample from a binomial distribution with
\(T_{r,t}^{s,\text{HIV}}\) trials each with a probability of
\(\rho_{r}(t)\):

\begin{equation}
P^{s,\text{HIV}}_{r,t}\sim\text{Binom}(T_{r,t}^{s,\text{HIV}}, \rho_{r}(t))
\end{equation}

These surveys are the most reliable data source we have for HIV
prevalence, but they are expensive and relatively infrequent. We might
have two or three surveys over the past 10 years in any given
sub-Saharan African nation.

On the other hand, ANC clinics report data regularly but are attended
exclusively by pregnant women, who we might expect to be at differential
risk of HIV infection relative to the general population. Therefore, we
cannot estimate \(\rho_{r}(t)\) with
\(P^{s,\text{HIV}}_{r,t}/T^{s,\text{HIV}}_{r,t}\) with. Instead,
following Bao (\protect\hyperlink{ref-bao_new_2012}{2012}), We estimate
site-specific ANC prevalence as a function of general population
prevalence and clinic effects:

\begin{equation}
\begin{split}
\text{logit}\;\rho^s_{r}(t) &= \text{logit}\;\rho_{r}(t) + \delta_s \\
\delta_s &\sim \text{N}(0, \sigma^2_\delta) \\
\sigma^2_s &\sim\text{N}^+(0,1)
\end{split}
\end{equation}

where \(\delta_s\) is a clinic-specific intercept. Then we can proceed
as above:

\begin{equation}
P^{s,\text{HIV}}_{r,t}\sim\text{Binom}(T_{r,t}^{s,\text{HIV}}, \rho^s_{r}(t)).
\end{equation}

ANC clinic HIV test results are less representative and more susceptible
to collection error than survey data, but they are available at far
greater temporal frequency.

\hypertarget{treatment}{%
\subsubsection{Treatment}\label{treatment}}

Several of the most recent household surveys conducted assays for the
presence of antiretroviral treatment (ART) in HIV-positive respondents'
blood samples.(ICAP at Columbia University and PEPFAR
\protect\hyperlink{ref-icap_at_columbia_university_methodology_2019}{2019})
Let \(T_{r,t}^{s,\text{ART}}\) represent the number of people tested for
ART, and let the \(P_{r,t}^{s,\text{HIV}}\) be the number of positive
tests. As above, we assume that these seroprevalence data provide an
unbiased estimate of true ART coverage in population segment
\(\{r,t\}\), denoted \(\alpha_{r}(t)\). Then we can make the same
assumption as before:

\begin{equation}
P^{s,\text{ART}}_{r,t}\sim\text{Binom}(T_{r,t}^{s,\text{ART}}, \alpha_{r}(t)).
\end{equation}

\hypertarget{programmatic-patient-count-data}{%
\subsubsection{Programmatic Patient Count
Data}\label{programmatic-patient-count-data}}

Because government-run clinics represent the largest administrators of
ART in most high-prevalence areas, we have an additional source of data
on ART coverage: facility-level ART patient counts. For a given
facility, we have the number of ART patients that facility treated over
a given time span (often quarterly intervals), which we will denote
\(C^s_{r,t}\). For convenience, we are currently using counts aggregated
to the region level: \(C_{r,t}\).

Given that we cannot measure the denominators for these data and that
they represent a nearly complete count of the number of adults receiving
ART in segment a given region at a given time, we need to model them as
count data. If we are willing to assume that the share of people on ART
who receive treatment outside of government run clinics is negligibly
small, then we can assume that \(C_{r,t}\) is an unbiased measurement of
the true number of adults on treatment at time \(t\) in region \(r\),
denoted \(A_r(t)\).

\hypertarget{a-likelihood-for-varyingpopulation-sizes}{%
\paragraph{A Likelihood for Varying~Population
Sizes}\label{a-likelihood-for-varyingpopulation-sizes}}

Fitting a model to these counts is a trickier problem that it might
seem. All else being equal, we would expect a large urban region to
serve more ART patients than a small rural region, so we need a model
that will make assumptions about variance that can work across regions
of varying population sizes.

Following Lindén and Mäntyniemi
(\protect\hyperlink{ref-linden_using_2011}{2011}), we have implemented a
negative binomial model with a variance that can scale both linearly and
quadratically with its mean. Say we are using the standard negative
binomial parameterization:

\begin{equation}
Pr(X= x;r,p) = \frac{\Gamma(x+r)}{x!\Gamma(r)}p^r(1-p)^x.
\end{equation}

Then \(\mathbb E(X) =\mu = r(1-p)/p\) and
\(\text{Var}(X) = \sigma^2 = r(1-p)/p^2\). We can solve for \(r\) and
\(p\) to see that \(r = \mu^2/(\sigma^2-\mu)\) and \(p = \mu/\sigma^2\).
In a traditional negative binomial model with \(\mu<\sigma^2\), we
estimate an overdispersion parameter, \(\theta > 0\), and define

\begin{equation}
\sigma^2 = \mu + \theta\mu^2.
\end{equation}

As \(\theta\to0\), \(\sigma^2 \to \mu\) and the distribution converges
to a Poisson distribution with rate parameter \(\mu\).

With this formulation, we can see why fitting a model simultaneously to
regions of varying size might be difficult. A single value of \(\theta\)
will impact regions of different sizes in radically different ways. A
lower value of \(\theta\) might explain the variation in a large region
well, while not adequately accounting for overdispersion in a small
region.

To help ease these problems, we use the formulation from Lindén and
Mäntyniemi (\protect\hyperlink{ref-linden_using_2011}{2011}):

\begin{equation}
\sigma^2 = \mu+ \omega\mu + \theta\mu^2,
\end{equation}

where \(\theta,\omega >0\). The addition of a linear term should help
avoid larger regions blowing smaller regions out of the water, so to
speak. Given fixed \(\theta\) and \(\mu\), as \(\omega \to 0\), this
distribution converges to a traditional negative binomial with
overdispersion \(\theta\). Conversely, given fixed \(\omega\) and
\(\mu\), as \(\theta\to0\), it converges to a ``quasi-Poisson''
distribution. We will write this three-parameter version of the negative
binomial distribution as \(\text{NegBinom}(\mu, \omega, \theta)\)
assuming that \(\sigma^2\) (and consequently \(r\) and \(p\)) is
calculated internally.

\hypertarget{cross-region-treatment-seeking}{%
\paragraph{Cross-Region Treatment
Seeking}\label{cross-region-treatment-seeking}}

The other critical problem we need to address when dealing with these
data are that they are collected from facilities, not households.
Patients can (and do) seek treatment outside of their
regions-of-residence (or ``home regions''), so we need to consider that
the observed patient count series for a given region is composed of
patients from all sufficiently ``close'' regions. For now, we will say
any two regions \(j\) and \(r\) are sufficiently close if they are
within some fixed degree of adjacency of each other (the number of
borders we need to cross to get from \(r\) to \(j\) is less than \(D\)).
The results presented below were generated by a model with \(D = 2\).
Let \(j\sim r\) denote \(j\) and \(r\) being adjacent, and let
\(\{k\sim r\}\) be the set of all regions that are ``close'' to \(r\).

We can define a multinomial model for the odds that an individual with
home region \(r\) will seek treatment in adjacent district \(j\) over
region \(r\):

\begin{equation}
\begin{split}
\text{log}\;\frac{P(A_{r\to j})}{P(A_{r\to r})} &= \frac{\text{log}\;m_j}{d_{i\to j}^2} \\
\text{log}\;m_j &\sim \text{N}(\text{log}\;\bar m_j, \sigma_m) \\
\sigma_m&\sim\text{N}^+(0, 1) \\
\bar m_j &= \frac{0.05}{\|\{k\sim j\}\setminus j\|}
\end{split}
\end{equation}

where \(m_j\) is a destination-specific intercept and \(d_{i\to j}\) is
the number of borders needed to cross to get from \(i\) to \(j\). Note
that \(P(A_{r\to r})/P(A_{r\to r}) =m_r= 1\), so we do not need a model
for each home region. We are currently defining the prior mean for each
\(m_j\) as \(0.05\) divided by the number of regions for which \(j\) is
a neighbor within \(D\) degrees, which is exactly
\(\{k\sim j\}\setminus j\). Roughly speaking, this means that we expect
5\% of ART patients in any region \(r\) to seek treatment outside of
\(r\). For all \(j\sim r\) (including \(r\) itself), we can find

\begin{equation}
P(A_{r\to j}) = \pi_{r\to j}= \frac{{m_j}}{\sum_{l\in\{k\sim r\} } {m_l}}.
\end{equation}

With \(\pi_{r \to j}\), we can allocate the total count of people
receiving treatment in \(r\) to each \(j\sim r\). We can estimate the
total number of people receiving treatment in \(r\) as

\begin{equation}
A^\star_r(t) = \sum_{j\in\{k\sim r\}} \pi_{j\to r}A_j(t).
\end{equation}

Note that we are using \(\pi_{j\to r}\) inside the summand, not
\(\pi_{r \to j}\); we are essentially collecting all of the ART patients
in \(j\) we believe are going from \(j\) to \(r\) to find the total
number of patients receiving treatment in \(r\).

At long last, we can define a likelihood for the ART patient count data.
Using the three-parameter negative binomial parameterisation from
before, we have

\begin{equation}
\begin{split}
C_{r,t}&\sim\text{NegBinom}(A^\star_{r}(t), \omega, \theta) \\
\text{log}\;\omega&\sim \text{N}(0, 2) \\
\text{log}\;\theta&\sim \text{N}(0, 2),
\end{split}
\end{equation}

where \(\text{log}\;\omega\) and \(\text{log}\;\theta\) are estimated
parameters. The priors on \(\omega\) and \(\theta\) are essentially
arbitrary and could certainly be improved. In a sense, we are fitting a
multinomial model on a matrix of flow counts for which we only observe
one set of margins.

\hypertarget{recency}{%
\subsubsection{Recency}\label{recency}}

Our final source of data is recency assays from the most recent wave of
household surveys (those conducted as a part of the PHIA program). These
data offer estimates of the proportion of people living with HIV (PLHIV)
who were infected in the past year, the closest thing to direct
incidence measurement available. Reusing the notation from the other
seroprevalence data sources, we have \(P_{r,t}^{s,\text{Rec}}\) and
\(T_{r,t}^{s,\text{Rec}}\). If we know the HIV incidence and prevalence
rates in a given demographic segment, \(\lambda_{r}(t)\) and
\(\rho_{r}(t)\) respectively, we can use the estimator from Kassanjee,
McWalter, and Welte (\protect\hyperlink{ref-kassanjee_short_2014}{2014})
to find the implied proportion of infections that should be recent:

\begin{equation}
\nu_{r}(t) = \frac{\lambda_{r}(t)\cdot(1-\rho_{r}(t))\cdot(\Omega_R-\gamma_R)+\gamma_R\rho_{r}(t)}{\rho_{r}(t)},\\
\end{equation}

where \(\Omega_R\) is the mean duration of recent infection (fixed at
130/365), and \(\gamma_R\) is the proportion of positive recency assays
that are false positives (fixed at 0).

As before, we assume that each \(P_{r,t}^{s,\text{Rec}}\) is a sample
from a binomial distribution:

\begin{equation}
P^{s,\text{Rec}}_{r,t}\sim\text{Binom}(T_{r,t}^{s,\text{Rec}}, \nu_{r}(t)).
\end{equation}

Although these data are the closest thing we have to direct measurement
of population-level incidence, they typically do not contain enough
information to be useful. For example, the recent MPHIA survey in Malawi
returned at least one positive recency assays in only nine of Malawi's
28 districts.

\hypertarget{inference-problem}{%
\subsection{Inference Problem}\label{inference-problem}}

As we have seen, none of the available data measure incidence directly,
but they are all generated from the same global HIV epidemic. Therefore,
if we can construct a model that estimates incidence and finds the
implied values of the measurable indicators, we can infer incidence.
More formally, say we are interested in estimation for a set of regions,
\(\{r_1, .. ., r_R\}\) and extent of time \([t_1, t_T]\). For
convenience, let \(r\) and \(t\) be arbitrary elements in
\(\{r_1, .. .,r_R\}\) and \([t_1,t_T]\), respectively. Then, our goal is
to infer a matrix \(\Lambda\) where \(\Lambda_{(r,t)}=\lambda_{r}(t)\)
given all available, relevant data \(\mathcal{D}\); that is, we want to
estimate \(P(\Lambda\;|\; \mathcal{D})\) where \(\mathcal{D}\) is the
collation of all indicators described in the previous section. Keeping
with previous literature (Brown et al.
\protect\hyperlink{ref-brown_improvements_2014}{2014}, for example), our
epidemic model will be deterministic, so we know that
\(P(\Lambda \;|\; \mathcal{D}) = P(\theta \;|\; \mathcal{D})\).
Therefore, we can use classical Bayesian inference:

\begin{equation}
P(\theta\;|\;\mathcal{D}) \propto P(\mathcal{D} \;|\; \theta) P(\theta).
\end{equation}

Within this framework, we need to define two components: a likelihood
\(P(\mathcal{D}\;|\;\theta)\) and a prior distribution \(P(\theta)\). We
have already defined the likelihood and priors for the observation model
with the distributional assumptions described above. In the next
section, we will describe the mechanistic model needed to relate
\(\Lambda\) to \(\mathcal{D}\) and the priors required to identify that
model.

\hypertarget{epidemic-model}{%
\subsection{Epidemic Model}\label{epidemic-model}}

We have made a set of distributional assumptions about the relationships
between three data sources (surveys, surveillance, and programmatic
counts) and the true, underlying HIV epidemic. To actually calculate
\(P(\mathcal{D};\theta^\star)\) for a given candidate set of parameters
\(\theta^\star\) we need to estimate the underlying epidemic.
Specifically, we need estimates of, prevalence, ART coverage, ART
patient counts, and incidence (\(\rho_r(t)\), \(\alpha_r(t)\),
\(A_r(t)\), and \(\lambda_r(t)\), respectively).

\hypertarget{compartmental-models}{%
\subsubsection{Compartmental Models}\label{compartmental-models}}

Compartmental models give us a way to build a generative model of the
disease indicators we need. Our model is essentially a variation of the
classical susceptible-infectious-recovered (SIR) model of infectious
disease. We track the populations in several mutually exclusive and
comprehensive compartments and define a system of ordinary differential
equations (ODEs) to govern rates of movement from one compartment to
another.

For example, the classical SIR model measures the number of susceptible,
infectious, and recovered individuals (\(S(t)\), \(I(t)\), and \(R(t)\),
respectively) in a closed population. Let \(N(t) = S(t) + I(t) + R(t)\)
be the size of the population at time \(t\). We can assume that
infectious individuals move from \(I\) to \(R\) according to some
recovery rate \(\gamma \in \mathbb{R}^+\). Further, if we assume the
principle of mass action holds, we can assume that individuals move from
\(S\) to \(I\) in proportion to the size of \(I\) according to a
transmission rate \(\kappa\in \mathbb{R}^+\). Then we can define the
whole model:

\begin{equation}
\begin{split}
\frac{\partial S(t)}{\partial t} &= -\kappa S(t)I(t)/N(t)\\
\frac{\partial I(t)}{\partial t} &=  \kappa S(t) I(t) /N(t) - \gamma I(t)\\
\frac{\partial R(t)}{\partial t} &= \gamma I(t).
\end{split}
\end{equation}

Fixing \(\kappa\), \(\gamma\), and the initial state of the system,
\((S(0), I(0), R(0))\), we can use the numerical method of our choosing
to find \((S(t), I(t), R(t))\) for any \(t > 0\).

In keeping with other work in this area, we are using the forward Euler
method, meaning we will be discretising the domain of the system of ODEs
(in this case, time) into intervals of length \(h\):

\begin{equation}
\begin{split}
S(t+h) &= S(t) + h\cdot(-\kappa S(t)I(t))\\
I(t+h) &= I(t) + h\cdot(\kappa S(t) I(t) - \gamma I(t))\\
R(t+h) &= R(t) + h\cdot(\gamma I(t)).
\end{split}
\end{equation}

If we define \(h\) to be some percentage of a calendar year, we can view
this as a discrete-time SIR model where \(h\) scales per-person-year
rates to a timescale of our choice.

\hypertarget{epidemic-model-of-hiv}{%
\subsubsection{Epidemic Model of HIV}\label{epidemic-model-of-hiv}}

\hypertarget{baseline-model}{%
\paragraph{Baseline Model}\label{baseline-model}}

For many reasons, this simple SIR model is not suited for HIV. We will
outline a few discrepancies here:

\begin{enumerate}
\def\labelenumi{\arabic{enumi}.}
\tightlist
\item
  Because HIV is a lifelong infection, the size of the \(R\) compartment
  is zero everywhere (ignoring two remarkable cases (Hütter et al.
  \protect\hyperlink{ref-hutter_long-term_2009}{2009}; Gupta et al.
  \protect\hyperlink{ref-gupta_hiv-1_2019}{2019})), and so we can drop
  it.
\item
  HIV prevalence rates can range far above 25\% in some populations
  (``HIV and AIDS in eSwatini. AVERT''
  \protect\hyperlink{ref-noauthor_hiv_2015}{2015}), so we cannot ignore
  demographic dynamics (non-HIV mortality, migration, and ageing-in to
  the adult population).
\item
  Untreated HIV is highly fatal, so we need to account for AIDS-related
  mortality (Yiannoutsos et al.
  \protect\hyperlink{ref-yiannoutsos_estimated_2012}{2012}; Todd et al.
  \protect\hyperlink{ref-todd_time_2007}{2007}; Marston et al.
  \protect\hyperlink{ref-marston_estimating_2007}{2007}).
\item
  Good adherence to ART drastically reduces infectiousness, so assuming
  that risk of infection is constant across contacts with all PLHIV
  regardless of treatment status will lead to biased estimates of
  incidence. Therefore, we need to add a compartment for people
  receiving treatment (Fonner et al.
  \protect\hyperlink{ref-fonner_effectiveness_2016}{2016}).
\end{enumerate}

Integrating these changes into the system of ODEs from before and
reformulating how we calculate incidence to reflect the prophylactic
effects of ART, we have the following:

\begin{equation}
\begin{split}
\frac{\partial S(t)}{\partial t} &= S(t)\cdot(-\lambda(t) - \mu^S) + E_t\\
\frac{\partial I(t)}{\partial t} &=  I(t)\cdot( - (\mu^S+\mu^I) - \alpha^\star_t) + \lambda(t)S(t) + \eta A(t)\\
\frac{\partial A(t)}{\partial t} &= A(t)\cdot(-\mu^S-\mu^A - \eta) + \alpha^\star_t I(t),
\end{split}
\end{equation}

where \(\mu^S\) is non-HIV mortality, \(E_t\) is the number of new
entrants to the population, \(\mu^I\) and \(\mu^A\) are HIV-related
mortality with and without ART, respectively, \(\alpha^\star_t\) is the
time-dependent rate at which PLHIV initiate ART, and \(\eta\) is the
rate of dropout from ART. We have rewritten incidence as
\(\lambda(t) = \kappa_t\rho(t)(1-\omega\alpha(t))\), where
\(\rho(t)=I(t)/N(t)\) is population prevalence, \(\omega \in [0,1]\) is
percent by which ART reduces infectiousness, and
\(\alpha(t) = A(t)/(I(t) + A(t))\) is ART coverage among all PLHIV.
Hence, \(\rho(t)(1-\omega\alpha(t))\) is ART-adjusted prevalence. We
have also allowed the HIV transmission rate to vary over time.

\hypertarget{incorporating-disease-progression}{%
\paragraph{Incorporating Disease
Progression}\label{incorporating-disease-progression}}

To more accurately reflect mortality patterns, we break the infected and
on-ART compartments into substages, which we will denote as \(I_c(t)\)
and \(A_c(t)\). Here \(c\) indexes something slightly different across
the two super-compartments. We track four substages for both \(I\) and
\(A\) defined by a single set of CD4 count intervals:
\(([0,200), [200,350), [350, 500), [500, \infty))\). We are currently
defining these compartments with CD4 count thresholds, but we could
equally define them with viral load thresholds. These specific
thresholds are borrowed from Johnson and Dorrington's Thembisa model
(Johnson and Dorrington
\protect\hyperlink{ref-johnson_thembisa_2019}{2019}).

For the \(I\) compartment, \(c\) indexes individuals' current CD4 count,
whereas for the \(A\) compartment, it indexes the individuals' CD4 count
at treatment initiation. We assume that individuals can ``move'' from
\(I_c\) to \(I_{c+1}\) but not from \(A_c\) to \(A_{c+1}\).

\begin{figure}

{\centering \includegraphics[width=1\linewidth]{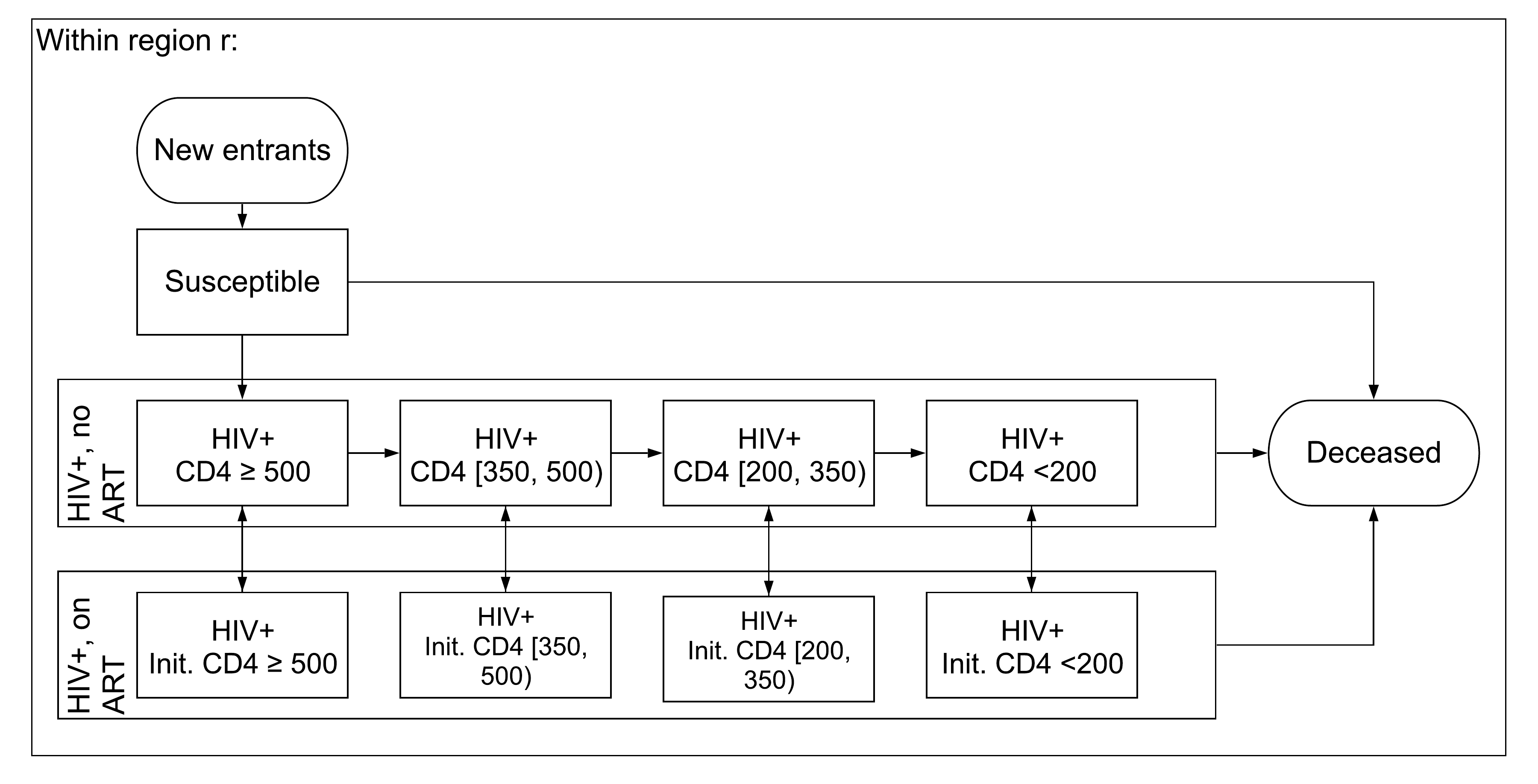} 

}

\caption{Diagram of our compartmental epidemic model of HIV focused on a single region}\label{fig:modelDiagram}
\end{figure}

Including these disease progression dynamics in the model, we get

\begin{equation}
\begin{aligned}
\frac{\partial S(t)}{\partial t} =& S(t)\cdot(-\lambda(t) - \mu^S) + E_t\\
\frac{\partial I_c(t)}{\partial t} =&  I_c(t)\cdot( - (\mu^S+\mu_c^I) - \alpha^\star_{c,t} - \tau_c) + \lambda_c(t)S(t) + \\
& \eta A_c(t) + \tau_{c-1}I_{c-1}(t) \\
\frac{\partial A_c(t)}{\partial t} =& A_c(t)\cdot(-\mu^S-\mu_c^A - \eta) + \alpha^\star_{c,t} I(t),
\end{aligned}
\end{equation}

where \(\tau_c\) is the rate of progression from disease stage \(c\) to
stage \(c+1\) for \(c\) less than the maximum value. We take our values
of \(\tau_c\) from Johnson and Dorrington
(\protect\hyperlink{ref-johnson_thembisa_2019}{2019}) Table 3.1. Note
that AIDS-related mortality now varies by disease stage. Figure
\ref{fig:modelDiagram} outlines the structure of this model.

\hypertarget{spatial-compartmental-model}{%
\paragraph{Spatial Compartmental
Model}\label{spatial-compartmental-model}}

Coming back to inferring incidence over space, we can construct a
(currently independent) model for each region \(r\) and arrive at our
final model:

\begin{equation}
\begin{aligned}
\frac{\partial S_r(t)}{\partial t} =& S_r(t)\cdot(-\lambda_r(t) - \mu^S) + E_{r,t}\\
\frac{\partial I_{r,c}(t)}{\partial t} =&  I_{r,c}(t)\cdot( - (\mu^S+\mu_c^I) - \alpha^\star_{r,c,t} - \tau_c) + \lambda_{r,c}(t)S(t) + \\ 
 & \eta A_{r,c}(t) + \tau_{c-1}I_{r,c-1}(t)\\
\frac{\partial A_{r,c}(t)}{\partial t} =& A_{r,c}(t)\cdot(-\mu^S-\mu_c^A - \eta) + \alpha^\star_{r,c,t} I(t).
\end{aligned}
\end{equation}

To project the HIV epidemic and find the indicators we need to evaluate
our likelihood, we need models for \(\lambda_r(t)\),
\(\alpha^\star_r(t)\), and \((S_r(0), I_{r,c}(0), A_{r,c}(0))\) for all
\(r\). Because we have not made any \emph{structural} changes to the
model, Figure \ref{fig:modelDiagram} is still accurate.

If the epidemic in each \(r\) were to be perfectly independent from that
of all other regions, our previous model of incidence would still be
valid. However, this assumption is demonstrably false in a model of
infectious disease. HIV first emerged in central Africa in the early
1900s and has since spread to every corner of the globe (Sharp and Hahn
\protect\hyperlink{ref-sharp_origins_2011}{2011}); clearly we cannot
claim that any two regions contain truly independent epidemics.

\begin{figure}

{\centering \includegraphics[width=1\linewidth]{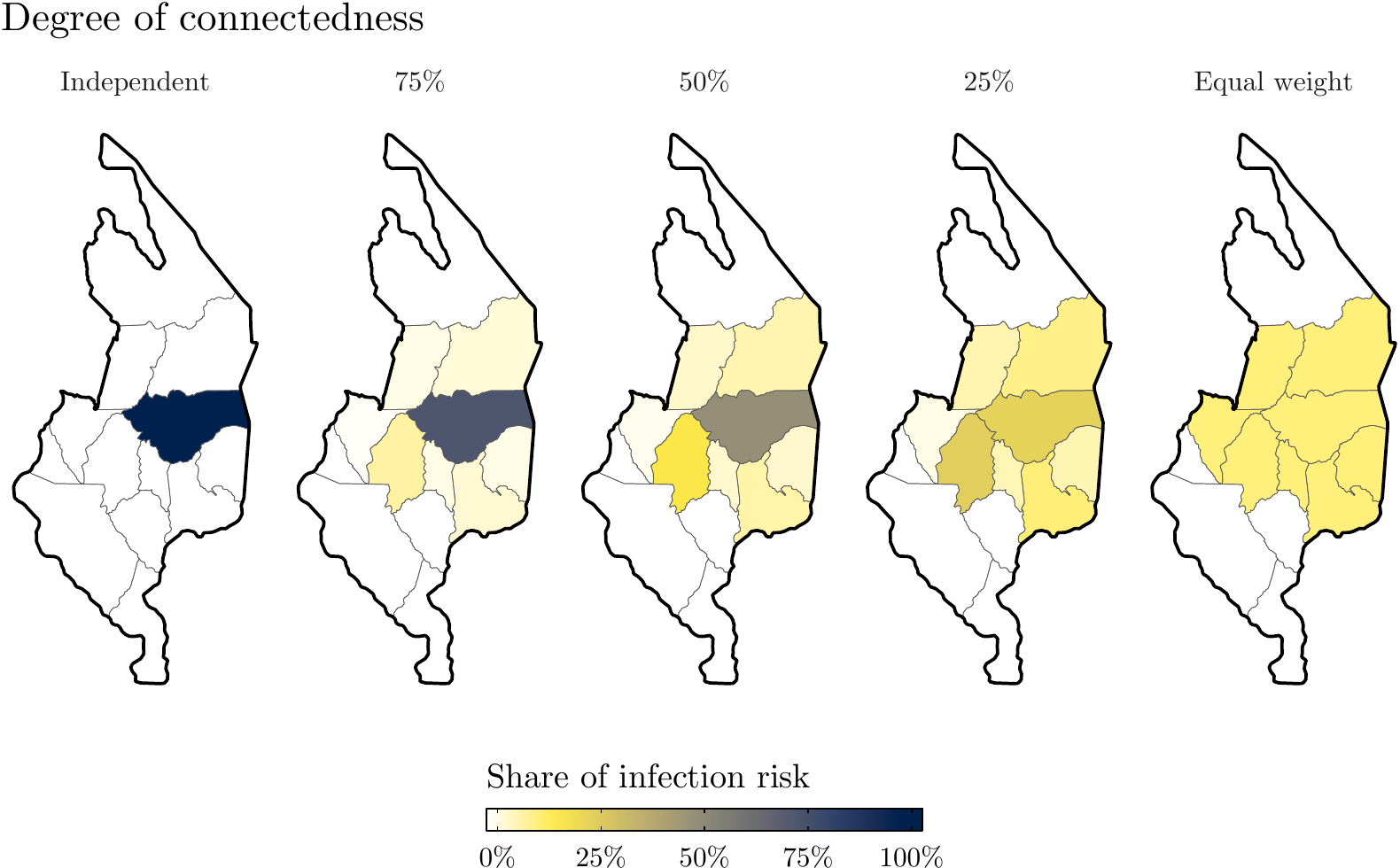} 

}

\caption{Values of $w(r,j)$ resulting from five assumptions about the degree of connectedness between districts of southern Malawi. Independent assumes that every infection originates internally, whereas "Equal weight" assumes an infection is as likely to come from any of Zomba's neighbours as from Zomba itself.}\label{fig:spatialConnectedness}
\end{figure}

We model the spread of disease over space directly by making incidence
in region \(r\) a function of prevalence in region \(r\) and all
adjacent regions (again, \(\{k\sim r\}\)). Specifically, we first model
the rate of infections attributable to disease stage \(c\) as

\begin{equation}
\log\lambda_{r,c}(t) = \log\kappa_{r,c,t} + \log\sum_{j\in\{k\sim r\}}w(r,j)\rho_{r,c}(t)(1-\omega\alpha_{r,c}(t)),
\end{equation}

where \(\kappa_{r,c,t}\) is a region-/substage-/time-specific
transmission rate, and \(w(r,j)\) is a weight proportionate to a measure
of distance between \(r\) and \(j\). Currently, we define \(w(r,j)\)
such that the share of risk coming from \(r\), \(w_0\), is fixed and the
remaining share is divided among its neighbors:

\begin{equation}
w(r,j) = \left\{\begin{aligned}
w_0 \qquad& \qquad r=j \\
\frac{p_j(1-w_0)}{\|\{j\sim r\} \setminus r\|}&\qquad r\neq j,
\end{aligned}\right.
\end{equation}

where \(p_j\) is the share of population at time 0 that lives in \(j\)
among neighbors of \(r\). Figure \ref{fig:spatialConnectedness}
illustrates five assumptions about the degree of spatial connectedness
between districts in the Southern region of Malawi relative to the Zomba
district.

We model the log-transformed region-/stage-/time-specific HIV
transmission rate \(\log \kappa_{r,c,t}\) as a hierarchical b-spline:
\begin{equation}
\begin{array}{rl}
\text{log}\;\kappa_{r,c,t} &= \log\xi_c +\sum_{i=1}^{K_\kappa+1}(\beta^\kappa_{i,0} + \beta^\kappa_{i,r})\phi_i(t) \\
\beta^\kappa_{i,0}&\sim \text{N}(0, 5) \\
\beta^\kappa_{i,r} &\sim \text{N}(0, \sigma_\kappa) \\
\beta^\kappa_{i,r}-\beta^\kappa_{i-1,r}&\sim\text{N}(0, 1) \\
\sigma_\kappa &\sim \text{N}^+(0, 1),
\end{array}
\end{equation}

where \(\xi_c\) is the relative infectiousness of stage \(c\),
\(K_\kappa\) is the number of knots, \(\phi\) is a b-spline basis
function of some order, \(\beta_{i,0}\) is a mean coefficient for basis
function \(i\), and \(\beta_{i,r}\) is a penalized region-specific
coefficient for region \(r\). Currently, we place knots at five-year
intervals and use a spline of order three. We use the relative
infectiousness ratios listed in Table 3.1 in Johnson and Dorrington
(\protect\hyperlink{ref-johnson_thembisa_2019}{2019}) to fix the values
of \(\xi_c\).

To obtain the HIV infection rate attributable to all disease stages
combined, we convert each \(\lambda_{r,c}(t)\) to a probability,
\(P(\text{inf}_{c};r,t) = 1-\exp(-h\cdot\lambda_{r,c}(t))\) assume
independence across disease stages, and aggregate as follows:

\begin{equation}
\begin{array}{rll}
P(\text{not inf}_{c};r,t)&=1-P(\text{inf}_{c};r,t) \\
&=\exp(-h\cdot\lambda_{r,c}(t)) \\
\\
P(\text{not inf};r,t)&=P(\cap_c\text{not inf}_{c};r,t) \\
&=\prod_{c}P(\text{not inf}_c;r,t) \\ 
&=\prod_c{\exp(-h \cdot\lambda_{r,c}(t))} \\
\\
P(\text{inf};r,t) &= 1-\prod_c\exp(-h\cdot \lambda_{r,c}(t)) \\
&=1-\exp(-h\sum\limits_c\lambda_{r,c}(t))
\end{array}
\end{equation}

Therefore, assuming the probability of infection from each disease stage
is independent from all other stages (that is, that
\(P(\text{not inf};r,t) =\prod_{c}P(\text{not inf}_c;r,t)\)), the
incidence rates are additive. Finally, we can find
\(\lambda_r(t) = 1/h \cdot \log P(\text{inf};r,t)\).

\hypertarget{model-of-art-initiation}{%
\paragraph{Model of ART Initiation}\label{model-of-art-initiation}}

Our model of the ART initiation rate among eligible disease stages is
similar to our model of incidence:

\begin{equation}
\begin{array}{rl}
\text{log}\;\alpha^\star_{r,c,t} &=\log \zeta_c + \sum_{i=1}^{K_\alpha+1}(\beta^{\alpha^\star}_{i,0} + \beta^{\alpha^\star}_{i,r})\phi_i(t) \\
\beta^{\alpha^\star}_{i,0}&\sim \text{N}(0, 5) \\
\beta^{\alpha^\star}_{i,r} &\sim \text{N}(0, \sigma_{\alpha^\star}) \\
\beta^{\alpha^\star}_{i,r}-\beta^{\alpha^\star}_{i-1,r}&\sim\text{N}(0, 1) \\
\sigma_{\alpha^\star} &\sim \text{N}^+(0, 1).
\end{array}
\end{equation}

Here, we place a knot every year and set \(\phi\) to be order one,
essentially making this model a random walk. For all \(t\) before ART
was scaled up in any region-of-interest, we fix \(\phi_i(t)\) to be
zero; additionally, we do not estimate any \(\beta^{\alpha^\star}_i\)
with support entirely before ART scale-up. The model used to generate
results presented on December 5th, 2019 fixed \(\zeta_c = 1\) for all
eligible \(c\), but the latest, less tested version sets
\(\zeta_c = \mu^I_c/\mu^I_1\); in other words, PLHIV at stage \(c\)
initiate treatment in proportion to the expected mortality in \(c\).

\hypertarget{estimation-of-initial-state}{%
\paragraph{Estimation of Initial
State}\label{estimation-of-initial-state}}

Our model can begin the epidemic projection at any point in time by
estimating the initial state of the compartmental model, denoted
\((S_r(0), I_{r,c}(0), A_{r,c}(0))\), with a small area model that draws
strength across regions. Specifically, we use logit-linear models to
estimate \(\rho_r(0)\) and \(\alpha_r(0)\) and use them to solve for
\((S_r(0), I_{r,c}(0), A_{r,c}(0))\).

Our model for initial prevalence is as follows:

\begin{equation}
\begin{split}
\text{logit}\;\rho_{r}(0) &= \rho_0 + \rho_r  \\
\rho_r &\sim \text{N}(0, \sigma_\rho) \\
\rho_0 &\sim \text{N}(0, 5) \\
\sigma_\rho &\sim \text{N}^+(0,1),
\end{split}
\end{equation}

where \(\rho_0\) is cross-region logit-transformed prevalence at time
\(0\) and \(\rho_r\) is a regional deviation from \(\rho_0\).

If time 0 is before ART scale-up, \(\alpha_r(0)\) is fixed to be zero in
all regions. Otherwise, we use an equivalent model:

\begin{equation}
\begin{split}
\text{logit}\;\alpha_{r}(0) &= \alpha_0 + \alpha_r  \\
\alpha_r &\sim \text{N}(0, \sigma_\alpha) \\
\alpha_0 &\sim \text{N}(0, 5) \\
\sigma_\alpha &\sim \text{N}^+(0,1).
\end{split}
\end{equation}

Making fixed assumptions about the share of each substage among people
with and without treatment, \(b^\alpha_c\) and \(b^\rho_c\),
respectively, we can find \(I_{r,c}(0)\), \(A_{r,c}(0)\), and
\(S_{r}(0)\):

\begin{equation}
\begin{array}{rl}
I_{r,c}(0) &=b^\rho_c \cdot(1-\alpha_{r}(0))\cdot\rho_{r}(0)\cdot P_{r}(0) \\
A_{r,c}(0) &=b^\alpha_c \cdot\alpha_{r}(0)\cdot\rho_{r}(0)\cdot P_{r}(0) \\
S_{r}(0) &= P_{r}(0) - \sum_c(I_{r,c}(0) + A_{r,c}(0)),
\end{array}
\end{equation}

where \(P_r(0)\) is exogenously defined population at time \(0\) and
\(b^\rho_c\) and \(b^\alpha_c\) are fixed assumptions about the share of
PLHIV without. Ideally, we would estimate \(b^\rho_c\) and
\(b^\alpha_c\), but too few sources provide CD4-specific data for that
to be possible. With a model for the initial state, we have everything
we need to project the epidemic using our compartmental model.

\hypertarget{implementation}{%
\subsection{Implementation}\label{implementation}}

We have implemented this model in C++ using the \texttt{TMB} R/C++
library (Kristensen et al.
\protect\hyperlink{ref-kristensen_tmb:_2016}{2016}). This software
allows users to write statistical models using the \texttt{Eigen} C++
library (Guennebaud, Jacob, and others
\protect\hyperlink{ref-guennebaud_eigen_2010}{2010}) and interact with
the compiled models through R. We used the sparse matrix tools built in
to \texttt{Eigen} to develop an efficient implementation of the Euler
method. Using automatic differentiation, \texttt{TMB} provides access to
the gradient functions of arbitrary statistical models and, as such, is
a powerful tool for inference and optimization. The inference strategy
built in to \texttt{TMB} optimizes the approximate marginal likelihood
with respect to a pre-defined set of ``random'' parameters and, after
finding the maximum a posterior (MAP) parameter estimates, assumes the
parameters form a multivariate normal distribution around those modes
(Kristensen et al. \protect\hyperlink{ref-kristensen_tmb:_2016}{2016};
Skaug and Fournier \protect\hyperlink{ref-skaug_automatic_2006}{2006}).

The TMB inference strategy makes extensive use of the Laplace
approximation and might therefore be poorly suited to produce samples
from distribution that is assymetric, bimodal, or otherwise non-normal.
This gap is filled by the \texttt{tmbstan} R package (Monnahan and
Kristensen \protect\hyperlink{ref-monnahan_no-u-turn_2018}{2018}), which
uses the objects that \texttt{TMB} builds to run the \texttt{rstan}
package's implementation of No-U-Turn sampler (NUTS) (Carpenter et al.
\protect\hyperlink{ref-carpenter_stan:_2017}{2017}). NUTS is a variant
of Hamiltonian Monte Carlo that can reliably produce samples from even
extremely complex statistical models and that we believe represents the
current gold standard for sampling algorithms. Through \texttt{tmbstan}
we can use either inference strategy. The results presented in this
paper and at Epidemics 7 were generated from posterior distributions
obtained through NUTS.

\hypertarget{application}{%
\subsection{Application}\label{application}}

We fit the model to district-level data from Malawi from the beginning
of 2000 to the end of 2018. Between January 1st, 2000 and December 31st,
2018, four nationally representative household surveys collected data on
HIV seroprevalence in Malawi: three DHS surveys (2004, 2010, and 2015)
(``The DHS Program - DHS Methodology''
\protect\hyperlink{ref-noauthor_dhs_nodate-1}{n.d.}) and one PHIA survey
(2015-2016) (ICAP at Columbia University and PEPFAR
\protect\hyperlink{ref-icap_at_columbia_university_methodology_2019}{2019}).
We used HIV seroprevalence data from each of these surveys, aggregating
test results to the district level. The MPHIA survey also provides
results of ART and recency blood tests. We incorporated HIV prevalence
data from sentinel ANC facilities, of which there are typically two per
district, using the hierarchical logistic model described above.
Finally, we used reported district-level ART patient counts to inform
the model of ART initiation.

\hypertarget{results}{%
\section{Results}\label{results}}

\begin{figure}

{\centering \includegraphics[width=1\linewidth]{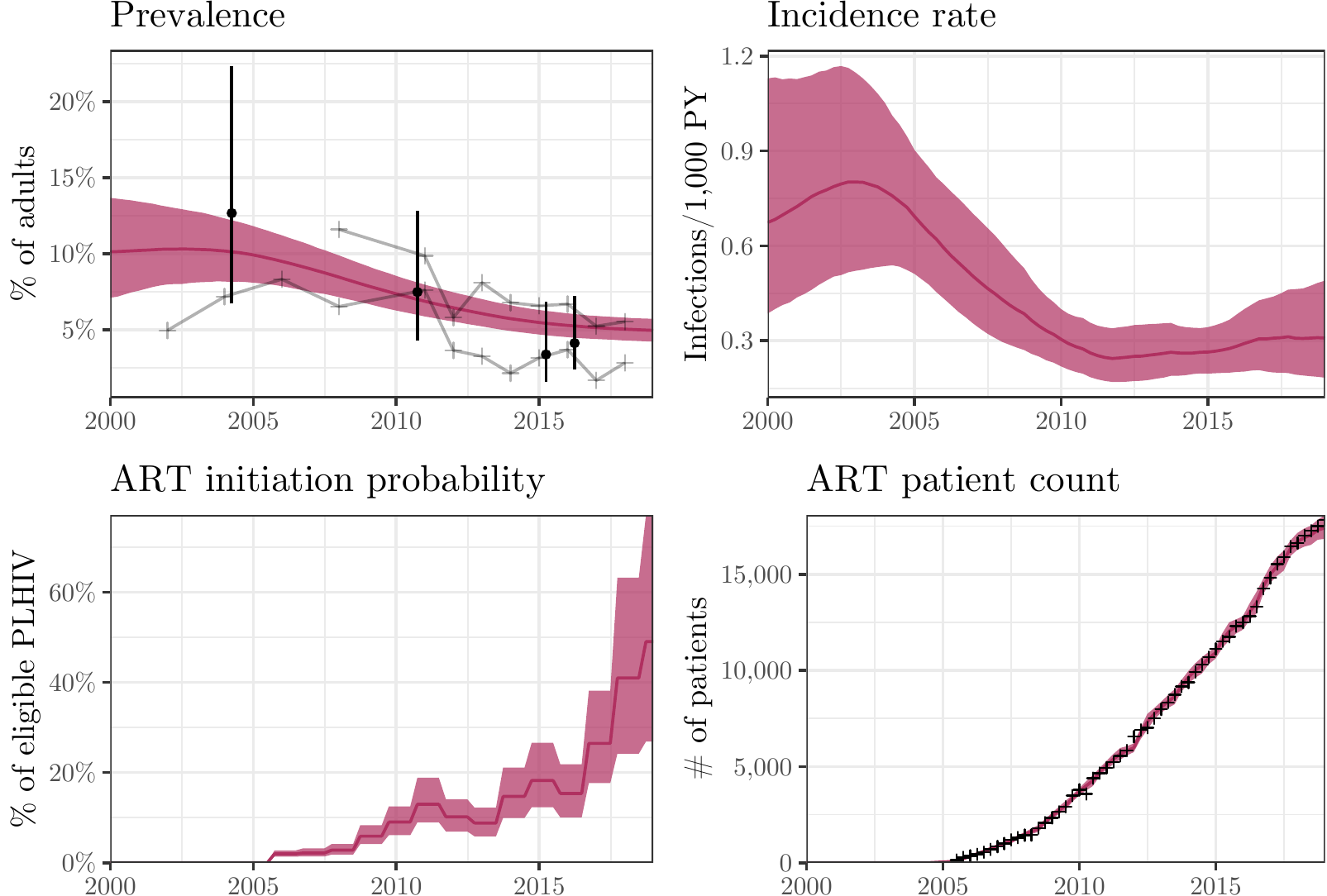} 

}

\caption{Data and estimates in the Dedza district of Malawi. Circular points and line  ranges represent point estimates and 95-percent confidence intervals from surveys. Connected series of crosses in the prevalence plot represent observed prevalence series from sentinel surveillance ANC clinics. Non-connected crosses in the ART patient count plot represent reported programmatic patient counts. Each red line is the median estimate for the  corresponding metric, and each red region is corresponding 95-percent uncertainty interval.}\label{fig:dedzaPlot}
\end{figure}

The model seems to perform well in applications to data from Malawi,
reconciling each of the included data sources without favoring one over
another. Figure \ref{fig:dedzaPlot} presents inferred incidence per
1,000 person-years and ART initiaton probabilities as well as fit to
data on prevalence and ART patient counts in the Dedza district of
Malawi. Because both incidence and underlying ART initiation rates are
difficult to measure directly, our ability to assess the validity of our
inference is limited. This hinderance is not unique to our study, but it
is worth highlighting.

\begin{figure}

{\centering \includegraphics[width=1\linewidth]{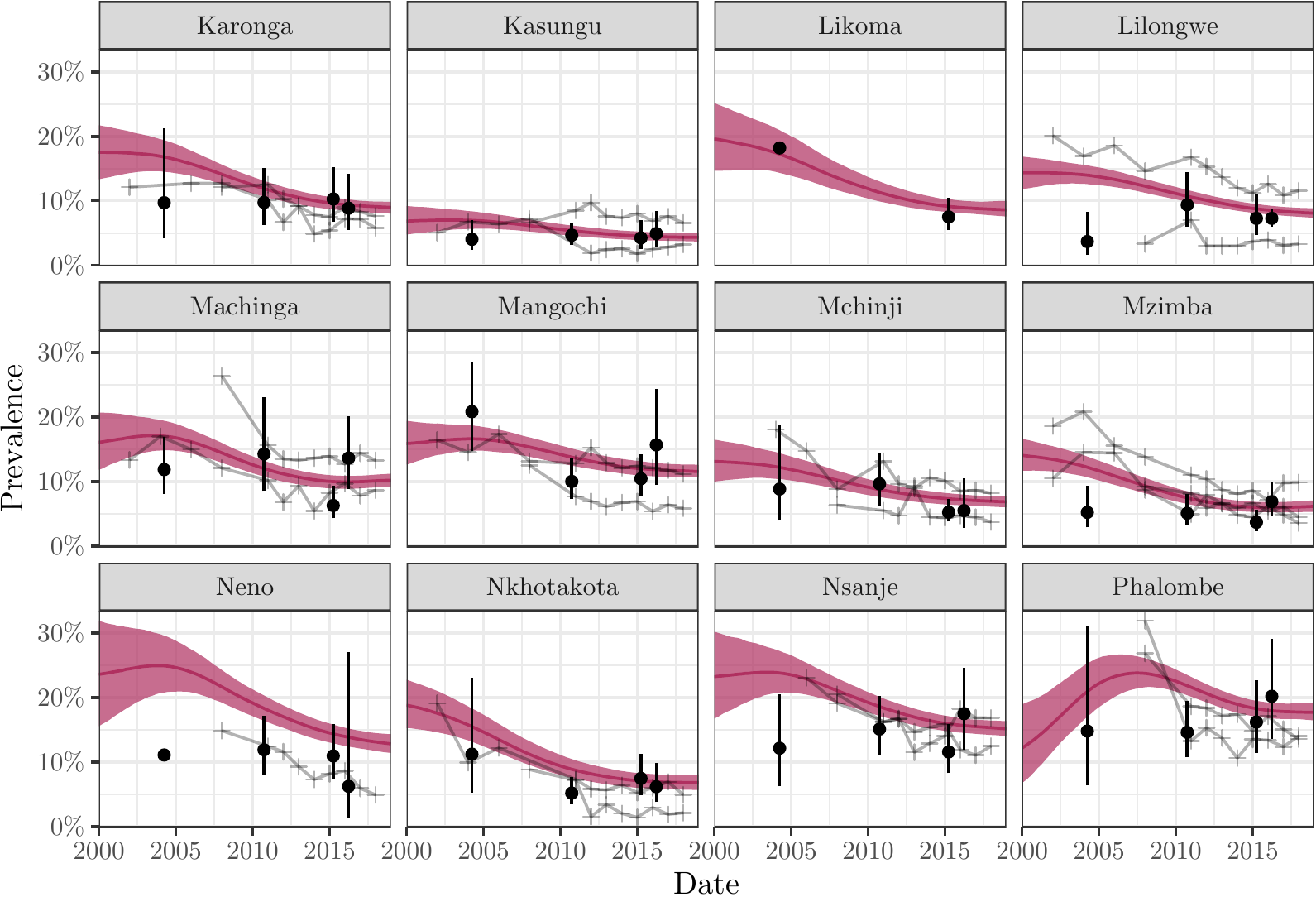} 

}

\caption{Data and estimates in 12 districts of Malawi. Circular points and line  ranges represent point estimates and 95-percent confidence intervals from surveys. Connected series of crosses in the prevalence plot represent observed prevalence series from sentinel surveillance ANC clinics. Each red line is the median estimate for the  corresponding metric, and each red region is corresponding 95-percent uncertainty interval.}\label{fig:allPrev}
\end{figure}

\begin{figure}

{\centering \includegraphics[width=1\linewidth]{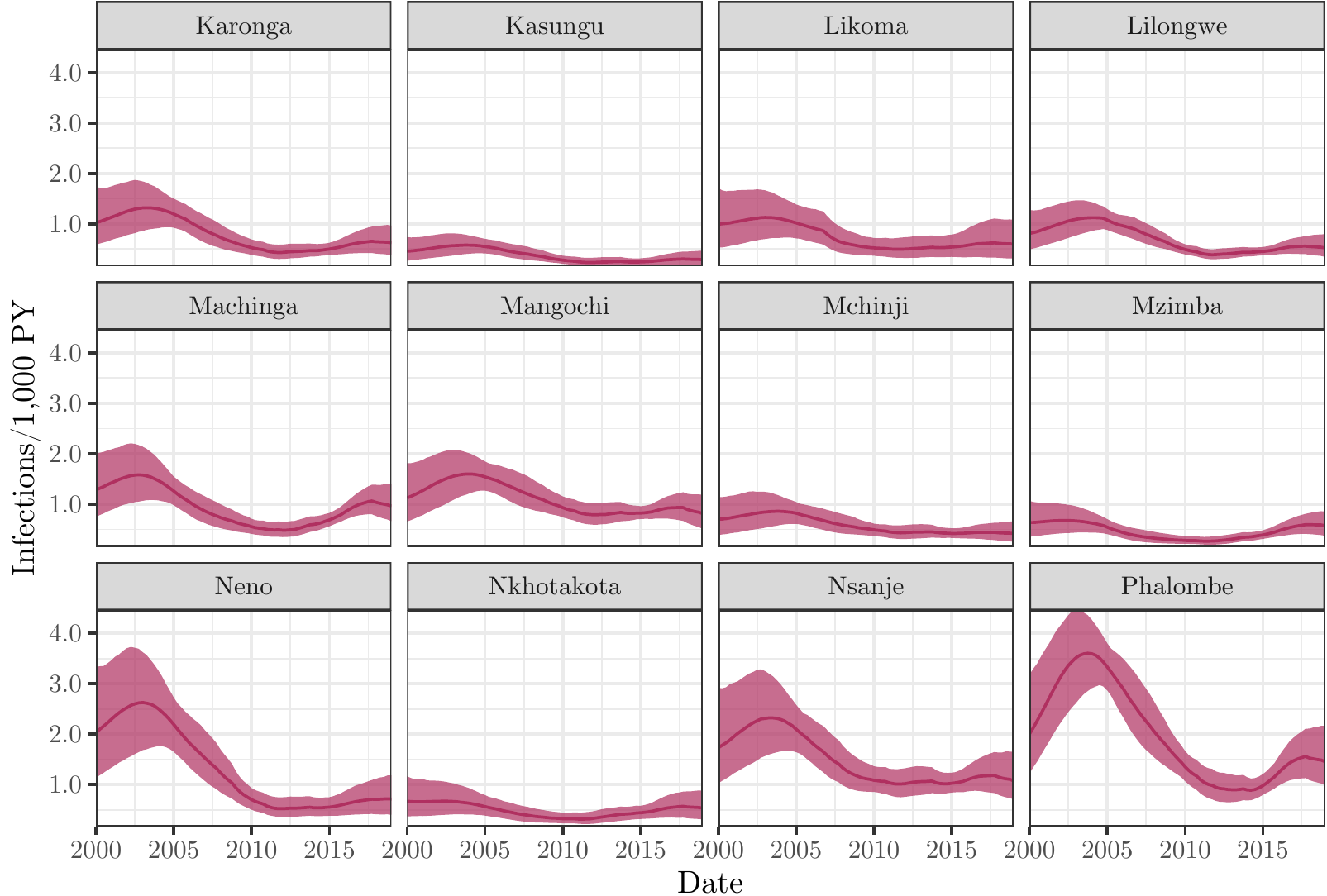} 

}

\caption{Data and estimates in 12 districts of Malawi. Each red line is the median estimate for the  corresponding metric, and each red region is corresponding 95-percent uncertainty interval.}\label{fig:allIncidence}
\end{figure}

\begin{figure}

{\centering \includegraphics[width=1\linewidth]{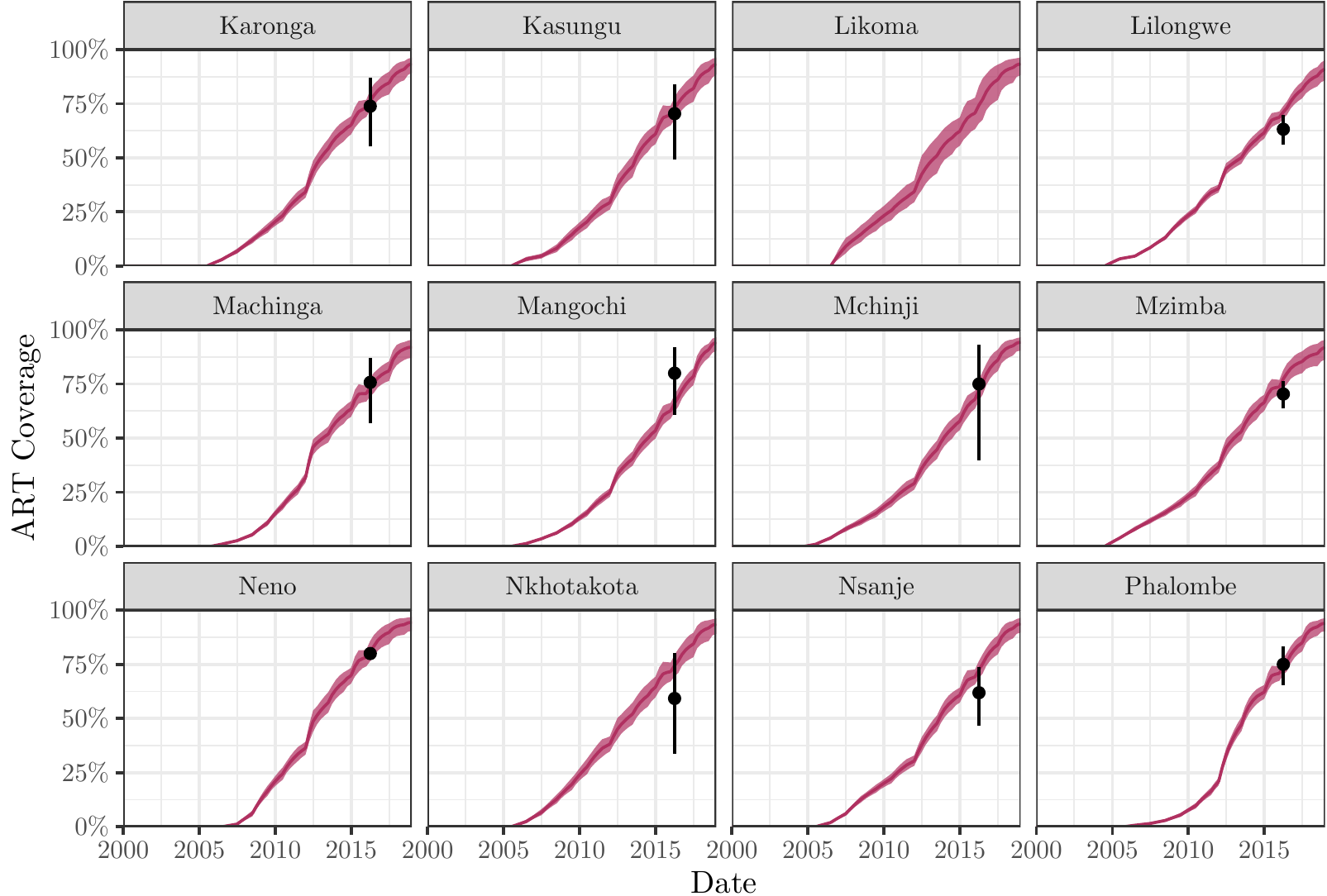} 

}

\caption{Data and estimates in 12 districts of Malawi. Circular points and line  ranges represent point estimates and 95-percent confidence intervals from surveys. Each red line is the median estimate for the  corresponding metric, and each red region is corresponding 95-percent uncertainty interval.}\label{fig:allARTCov}
\end{figure}

\begin{figure}

{\centering \includegraphics[width=1\linewidth]{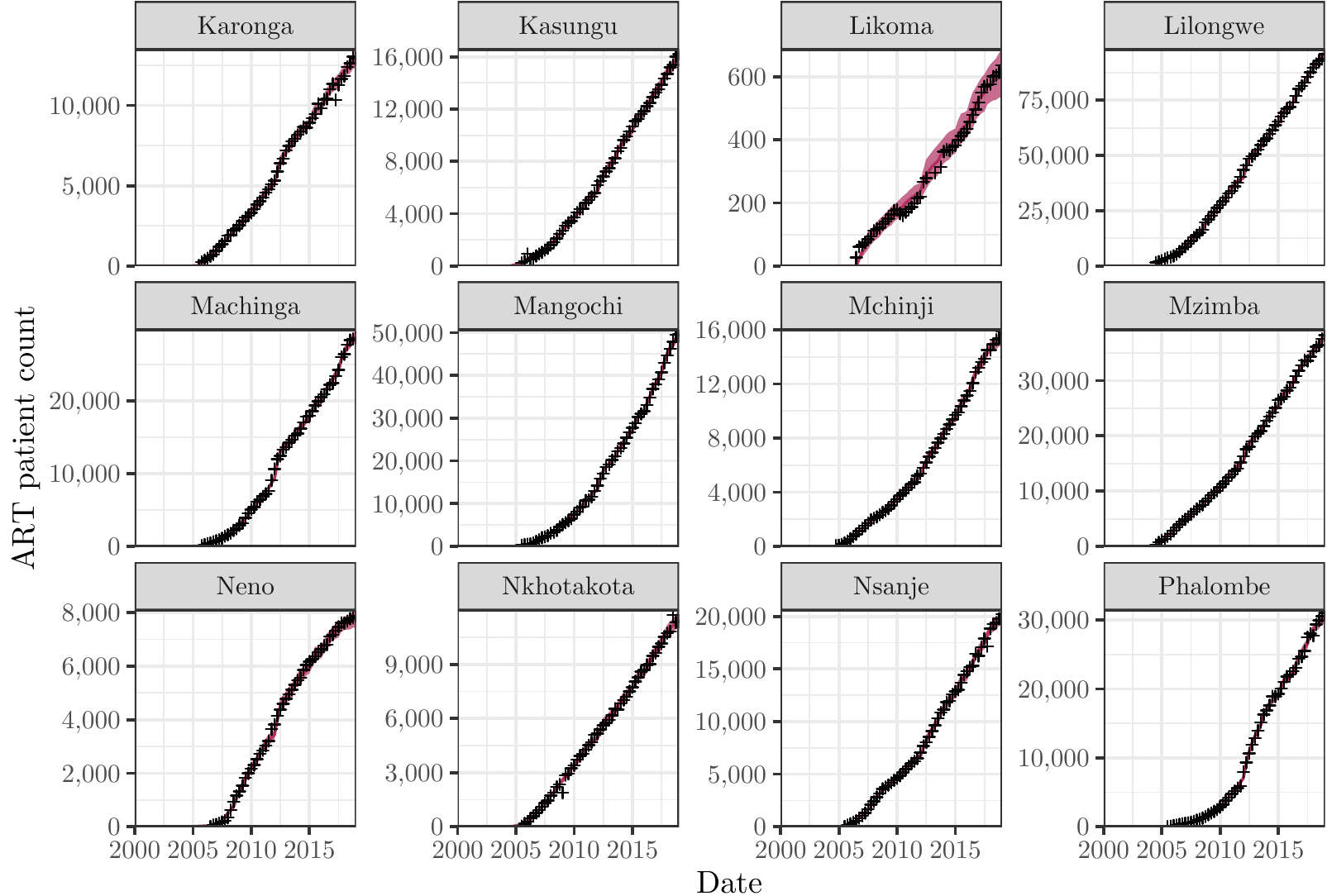} 

}

\caption{Data and estimates in 12 districts of Malawi. Non-connected crosses in the ART patient count plot represent reported programmatic patient counts. Each red line is the median estimate for the  corresponding metric, and each red region is corresponding 95-percent uncertainty interval.}\label{fig:allARTCount}
\end{figure}

Figures \ref{fig:allPrev}, \ref{fig:allIncidence}, \ref{fig:allARTCov},
and \ref{fig:allARTCount}, and present our estimates of prevalence,
incidence, ART coverage, and ART patient counts for 12 of Malawi's 28
districts. We specifically included Likoma and Phalombe and selected the
remaining 10 districts at random. To conserve space, we are only
including plots of these 12 districts. The full sets of plots are
readily available upon request.

Figure \ref{fig:allPrev} shows that the model seems to capture variation
in HIV prevalence over space, time, and their interaction reasonably
well. We are over estimating prevalence across the board in Neno, but
the survey estimates there are so uncertain that we cannot include them
on the plot. Interestingly, we predict that the epidemic peaked
significantly later in Phalombe than in other districts. We also note
that the model fit well to data from Likoma, for which we have far less
data than for any other district.

In Figure \ref{fig:allIncidence}, we see that our inferred incidence
series also exhibit considerable heterogeneity over space and time. For
example, the late prevalence peak we observed in Phalombe is reflected
in a later, higher peak incidence. This figure also illustrates the
value of fitting to multiple regions at once: we would never be able to
produce incidence estimates in Likoma if we did not share information
across regions.

\begin{figure}

{\centering \includegraphics[width=1\linewidth]{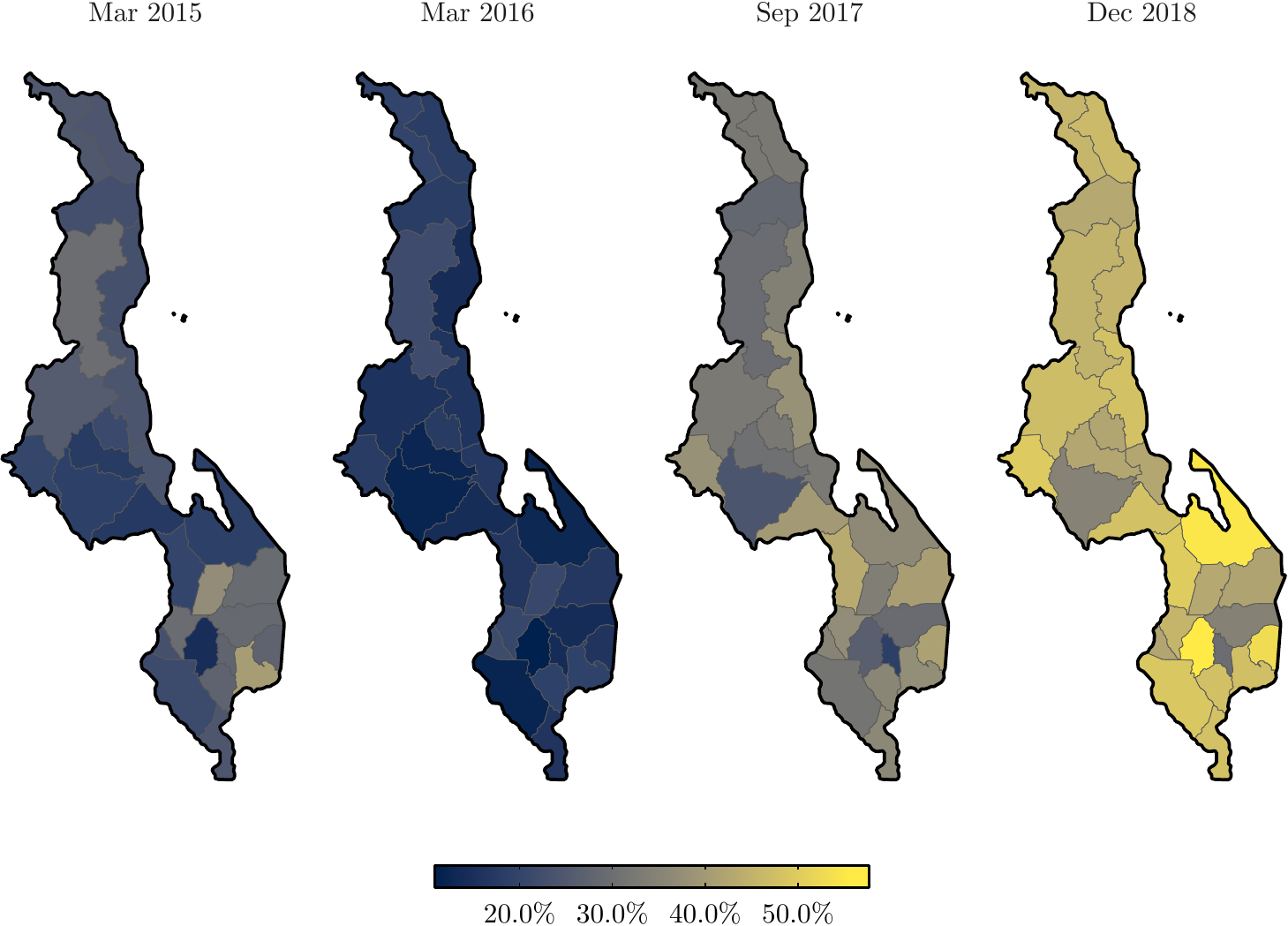} 

}

\caption{Inferred median ART initiation probabilities in Malawi, 2015-2018.}\label{fig:initMap}
\end{figure}

Taking Figures \ref{fig:allARTCov} and \ref{fig:allARTCount} as a pair,
we see that our model of ART initiation both produces plausible time
series of both ART coverage and ART patient counts. We recognize that
this is a particularly easy (exceptionally linear) case, but we believe
that the results are still encouraging. Further work is required to
thoroughly interrogate the model of ART initiation, but valid inference
of the spatio-temporal variation in ART initiation could be a
significant aid to policymakers. Figure \ref{fig:initMap} shows these
estimates for the current set of results from the beginning of 2015 to
the end of the study period. We can see that, even in the most recent
timepoint, there is considerable heterogeneity.

\begin{figure}

{\centering \includegraphics[width=1\linewidth]{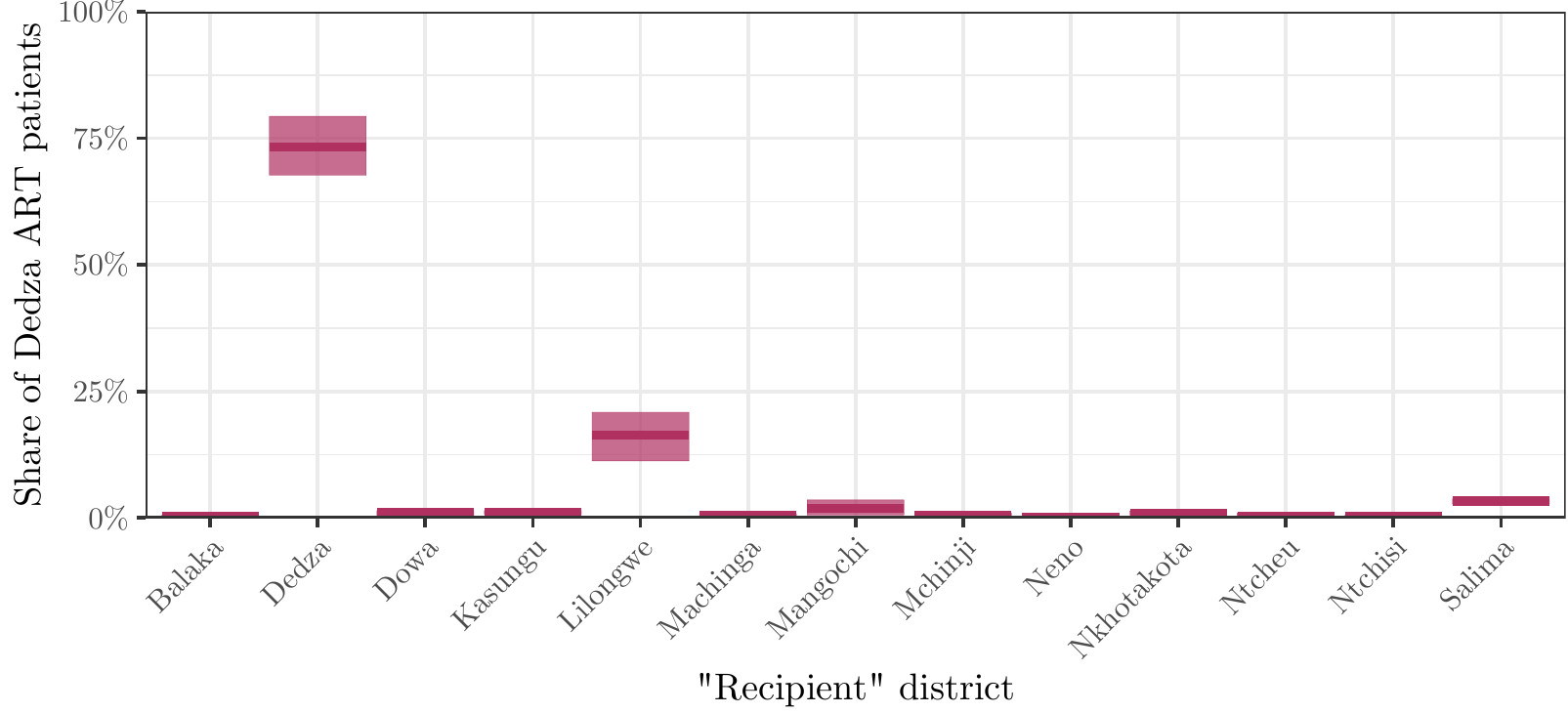} 

}

\caption{Estimated geographic distribution of treatment seeking among PLHIV residing in Dedza (median shares and 95-percent uncertainty intervals).}\label{fig:dedzaAlloc}
\end{figure}

Our model of cross-region treatment seeking is also difficult to
validate but seems to be performing well. Figure \ref{fig:dedzaAlloc}
shows the probability a person receiving ART in Dedza will seek
treatment each of the districts that are within two degrees of adjacency
from Dedza. The majority of treatment-seekers in Dedza stayed in their
home region, but a significant portion went to Lilongwe, which contains
the capital and largest city. Without allowing for cross-region
treatment seeking, our model consistently overestimates prevalence in
districts like Lilongwe and Chiradzulu, which have a longer history of
high-quality treatment provision.

Most importantly, the model fits well to each of the data sources we
provide to it. By adding flexibility in the models of HIV transmission
rates and ART initiation probabilities, we can fit simultaneously to
both traditional (seroprevalence measurements) and non-traditional
(programmatic counts) data sources.

\hypertarget{conclusion}{%
\section{Conclusion}\label{conclusion}}

In this study, we present a multivariate model of population-level HIV
that combines all available population-level data sources and exploits
the spatial structure of existing data in order to facilitate more
credible inference of HIV incidence at relatively granular geographic
resolutions. Incidentally, the supporting spatio-temporal models of HIV
transmission rates, ART initiation rates, and cross-region treatment
seeking provide estimates of politically relevant indicators that cannot
be obtained from the other models widely in use.

We believe that our work illustrates several key points:

\begin{enumerate}
\def\labelenumi{\arabic{enumi}.}
\tightlist
\item
  \textbf{Compartmental models of infectious disease are compatible with
  modern inferential frameworks.} Using sparse matrices, we were able to
  cut down the cost of integrating our model to the point where, even at
  relatively large geographic scales, we can run the inference procedure
  on a laptop. We suspect that further computational gains could be
  found by performing calculations in parallel or on a graphics
  processing unit.
\item
  \textbf{Accounting for the inherent spatial dynamics of HIV is
  critical at granular geographic resolutions.} We observed considerable
  spatio-temporal heterogeneity in incidence, transmission, ART
  initiation, and cross-region treatment seeking. Neglecting to account
  for any of these spatial dynamics would hamper our inference of the
  others. In particular, models that did not allow individuals to seek
  treatment outside of their home region severely overestimated
  prevalence (and, hence, incidence) in many regions. By letting the
  model ``reallocate'' patients to nearby districts, we allow it to
  explain high ART patient counts without increasing prevalence to
  implausibly high levels.
\item
  \textbf{The combined set of data sources is more valuable than the sum
  of its parts.} Our compartmental model is an explicit link between
  each of the various observable indicators, so every indicator informs
  our estimates of every other indicator. For example, a proposed
  parameter set that predicts very high HIV prevalence would result in
  poor fit to not only the direct observation of prevalence, but also to
  ART coverage and ART patient counts.
\item
  \textbf{Simultaneous modeling of multiple spatial units provides
  better inference in all cases.} First, shrinkage priors allow us to
  estimate the initial state of the epidemic across regions even if some
  of those regions are data-sparse. Additionally, giving the model the
  flexibility to explain observations in one area with epidemic patterns
  in another area helps avoid implausible estimates. For example,
  sustained high prevalence in one area might be the result of poor
  treatment coverage in a nearby area.
\end{enumerate}

\hypertarget{future-work}{%
\subsection{Future Work}\label{future-work}}

We are actively working on this model and foresee several improvements.

\begin{enumerate}
\def\labelenumi{\arabic{enumi}.}
\tightlist
\item
  \textbf{Greater demographic detail:} Incorporating age and sex
  dynamics will allow us to build a more realistic model of mortality
  and population.
\item
  \textbf{Incorporating population data:} The model currently includes a
  demographic projection, but we neither fit to nor match any data on
  population levels. We plan to explore a strategy where the total
  population in each time is fixed (from exogenous estimation processes)
  and migration is calculated as the difference between modeled
  population and observed population. Alternatively, we could attempt to
  fit to census data, but we suspect that that is out of scope for this
  particular project.
\item
  \textbf{Designing cross-validation scheme:} We are currently fitting
  to all available data, but we would like to devise a cross-validation
  scheme that prioritizes validity in more recent years. The challenge
  is that we have relatively little data to begin with.
\item
  \textbf{Tools for model comparison:} Because this model is complex and
  nonlinear, we do not have access to the typical set of model
  comparison tools we would use in a regression setting. Finding the
  appropriate metrics to assess the effects of changes to both the
  structure of the model and the inferential procedure will be an
  important step in interrogating our results.
\end{enumerate}

\hypertarget{acknowledgements}{%
\subsection{Acknowledgements}\label{acknowledgements}}

We are enormously grateful to our collaborators at the Department of HIV
\& AIDS in the Malawi Ministry of Health for their guidance and for
providing access to these data. We also thank the Imperial DIDE HIV
Inference Research Group for their helpful feedback.

Timothy M Wolock's work is funded through the Imperial President's PhD
Scholarship.

\newpage

\hypertarget{references}{%
\section*{References}\label{references}}
\addcontentsline{toc}{section}{References}

\hypertarget{refs}{}
\leavevmode\hypertarget{ref-bao_new_2012}{}%
Bao, Le. 2012. ``A New Infectious Disease Model for Estimating and
Projecting HIV/AIDS Epidemics.'' \emph{Sexually Transmitted Infections}
88 (Suppl\_2): i58--i64.
\url{https://doi.org/10.1136/sextrans-2012-050689}.

\leavevmode\hypertarget{ref-brown_improvements_2014}{}%
Brown, Tim, Le Bao, Jeffrey W. Eaton, Daniel R. Hogan, Mary Mahy,
Kimberly Marsh, Bradley M. Mathers, and Robert Puckett. 2014.
``Improvements in Prevalence Trend Fitting and Incidence Estimation in
EPP 2013.'' \emph{AIDS} 28 (November): S415.
\url{https://doi.org/10.1097/QAD.0000000000000454}.

\leavevmode\hypertarget{ref-carpenter_stan:_2017}{}%
Carpenter, Bob, Andrew Gelman, Matthew D. Hoffman, Daniel Lee, Ben
Goodrich, Michael Betancourt, Marcus Brubaker, Jiqiang Guo, Peter Li,
and Allen Riddell. 2017. ``Stan: A Probabilistic Programming Language.''
\emph{Journal of Statistical Software} 76 (1): 1--32.
\url{https://doi.org/10.18637/jss.v076.i01}.

\leavevmode\hypertarget{ref-cuadros_mapping_2017}{}%
Cuadros, Diego F., Jingjing Li, Adam J. Branscum, Adam Akullian, Peng
Jia, Elizabeth N. Mziray, and Frank Tanser. 2017. ``Mapping the Spatial
Variability of HIV Infection in Sub-Saharan Africa: Effective
Information for Localized HIV Prevention and Control.'' \emph{Scientific
Reports} 7 (1): 9093. \url{https://doi.org/10.1038/s41598-017-09464-y}.

\leavevmode\hypertarget{ref-dwyer-lindgren_mapping_2019}{}%
Dwyer-Lindgren, Laura, Michael A. Cork, Amber Sligar, Krista M. Steuben,
Kate F. Wilson, Naomi R. Provost, Benjamin K. Mayala, et al. 2019.
``Mapping HIV Prevalence in Sub-Saharan Africa Between 2000 and 2017.''
\emph{Nature}, May, 1. \url{https://doi.org/10.1038/s41586-019-1200-9}.

\leavevmode\hypertarget{ref-fonner_effectiveness_2016}{}%
Fonner, Virginia A., Sarah L. Dalglish, Caitlin E. Kennedy, Rachel
Baggaley, Kevin R. O'Reilly, Florence M. Koechlin, Michelle Rodolph,
Ioannis Hodges-Mameletzis, and Robert M. Grant. 2016. ``Effectiveness
and Safety of Oral HIV Preexposure Prophylaxis for All Populations.''
\emph{AIDS (London, England)} 30 (12): 1973--83.
\url{https://doi.org/10.1097/QAD.0000000000001145}.

\leavevmode\hypertarget{ref-guennebaud_eigen_2010}{}%
Guennebaud, Gaël, Benoı\^{}t Jacob, and others. 2010. \emph{Eigen V3}.
\url{http://eigen.tuxfamily.org}.

\leavevmode\hypertarget{ref-gupta_hiv-1_2019}{}%
Gupta, Ravindra K., Sultan Abdul-Jawad, Laura E. McCoy, Hoi Ping Mok,
Dimitra Peppa, Maria Salgado, Javier Martinez-Picado, et al. 2019.
``HIV-1 Remission Following CCR5Delta32/Delta32 Haematopoietic Stem-Cell
Transplantation.'' \emph{Nature} 568 (7751): 244--48.
\url{https://doi.org/10.1038/s41586-019-1027-4}.

\leavevmode\hypertarget{ref-gutreuter_improving_2019}{}%
Gutreuter, Steve, Ehimario Igumbor, Njeri Wabiri, Mitesh Desai, and
Lizette Durand. 2019. ``Improving Estimates of District HIV Prevalence
and Burden in South Africa Using Small Area Estimation Techniques.''
\emph{PLOS ONE} 14 (2): e0212445.
\url{https://doi.org/10.1371/journal.pone.0212445}.

\leavevmode\hypertarget{ref-hall_introduction_1997}{}%
Hall, D.L., and J. Llinas. 1997. ``An Introduction to Multisensor Data
Fusion.'' \emph{Proceedings of the IEEE} 85 (1): 6--23.
\url{https://doi.org/10.1109/5.554205}.

\leavevmode\hypertarget{ref-noauthor_hiv_2015}{}%
``HIV and AIDS in eSwatini. AVERT.'' 2015. July 21, 2015.
\url{https://www.avert.org/professionals/hiv-around-world/sub-saharan-africa/swaziland}.

\leavevmode\hypertarget{ref-hutter_long-term_2009}{}%
Hütter, Gero, Daniel Nowak, Maximilian Mossner, Susanne Ganepola, Arne
Müßig, Kristina Allers, Thomas Schneider, et al. 2009. ``Long-Term
Control of HIV by CCR5 Delta32/Delta32 Stem-Cell Transplantation.''
\emph{New England Journal of Medicine} 360 (7): 692--98.
\url{https://doi.org/10.1056/NEJMoa0802905}.

\leavevmode\hypertarget{ref-icap_at_columbia_university_methodology_2019}{}%
ICAP at Columbia University, and PEPFAR. 2019. ``Methodology. PHIA.''
2019. \url{https://phia.icap.columbia.edu/methodology/}.

\leavevmode\hypertarget{ref-johnson_thembisa_2019}{}%
Johnson, Leigh, and Rob Dorrington. 2019. \emph{Thembisa Version 4.2: A
Model for Evaluating the Impact of HIV/AIDS in South Africa}.
\url{https://www.thembisa.org/content/downloadPage/Thembisa4_2report}.

\leavevmode\hypertarget{ref-kassanjee_short_2014}{}%
Kassanjee, Reshma, Thomas A. McWalter, and Alex Welte. 2014. ``Short
Communication: Defining Optimality of a Test for Recent Infection for
HIV Incidence Surveillance.'' \emph{AIDS Research and Human
Retroviruses} 30 (1): 45--49.
\url{https://doi.org/10.1089/aid.2013.0113}.

\leavevmode\hypertarget{ref-kristensen_tmb:_2016}{}%
Kristensen, Kasper, Anders Nielsen, Casper W. Berg, Hans Skaug, and Brad
Bell. 2016. ``TMB: Automatic Differentiation and Laplace
Approximation.'' \emph{Journal of Statistical Software} 70 (5).
\url{https://doi.org/10.18637/jss.v070.i05}.

\leavevmode\hypertarget{ref-linden_using_2011}{}%
Lindén, Andreas, and Samu Mäntyniemi. 2011. ``Using the Negative
Binomial Distribution to Model Overdispersion in Ecological Count
Data.'' \emph{Ecology} 92 (7): 1414--21.
\url{https://doi.org/10.1890/10-1831.1}.

\leavevmode\hypertarget{ref-marston_estimating_2007}{}%
Marston, Milly, Jim Todd, Judith R. Glynn, Kenrad E. Nelson, Ram
Rangsin, Tom Lutalo, Mark Urassa, et al. 2007. ``Estimating `Net'
HIV-Related Mortality and the Importance of Background Mortality
Rates.'' \emph{AIDS (London, England)} 21 (Suppl 6): S65--S71.
\url{https://doi.org/10.1097/01.aids.0000299412.82893.62}.

\leavevmode\hypertarget{ref-meyer-rath_targeting_2018}{}%
Meyer-Rath, Gesine, Jessica B. McGillen, Diego F. Cuadros, Timothy B.
Hallett, Samir Bhatt, Njeri Wabiri, Frank Tanser, and Thomas Rehle.
2018. ``Targeting the Right Interventions to the Right People and
Places: The Role of Geospatial Analysis in HIV Program Planning.''
\emph{AIDS (London, England)} 32 (8): 957--63.
\url{https://doi.org/10.1097/QAD.0000000000001792}.

\leavevmode\hypertarget{ref-noauthor_monitoring_2017}{}%
``Monitoring, Evaluation, and Reporting (MER 2.0) Indicator Reference
Guide.'' 2017. PEPFAR.
\url{https://www.pepfar.gov/documents/organization/263233.pdf}.

\leavevmode\hypertarget{ref-monnahan_no-u-turn_2018}{}%
Monnahan, Cole C., and Kasper Kristensen. 2018. ``No-U-Turn Sampling for
Fast Bayesian Inference in ADMB and TMB: Introducing the Adnuts and
Tmbstan R Packages.'' \emph{PLoS ONE} 13 (5).
\url{https://doi.org/10.1371/journal.pone.0197954}.

\leavevmode\hypertarget{ref-sharp_origins_2011}{}%
Sharp, Paul M., and Beatrice H. Hahn. 2011. ``Origins of HIV and the
AIDS Pandemic.'' \emph{Cold Spring Harbor Perspectives in Medicine:} 1
(1). \url{https://doi.org/10.1101/cshperspect.a006841}.

\leavevmode\hypertarget{ref-skaug_automatic_2006}{}%
Skaug, Hans J., and David A. Fournier. 2006. ``Automatic Approximation
of the Marginal Likelihood in Non-Gaussian Hierarchical Models.''
\emph{Computational Statistics \& Data Analysis} 51 (2): 699--709.
\url{https://doi.org/10.1016/j.csda.2006.03.005}.

\leavevmode\hypertarget{ref-stover_updates_2017}{}%
Stover, John, Tim Brown, Robert Puckett, and Wiwat Peerapatanapokin.
2017. ``Updates to the Spectrum/Estimations and Projections Package
Model for Estimating Trends and Current Values for Key HIV Indicators.''
\emph{AIDS} 31 (April): S5.
\url{https://doi.org/10.1097/QAD.0000000000001322}.

\leavevmode\hypertarget{ref-noauthor_dhs_nodate-1}{}%
``The DHS Program - DHS Methodology.'' n.d. Accessed June 3, 2019.
\url{https://dhsprogram.com/What-We-Do/Survey-Types/DHS-Methodology.cfm}.

\leavevmode\hypertarget{ref-todd_time_2007}{}%
Todd, Jim, Judith R. Glynn, Milly Marston, Tom Lutalo, Sam Biraro,
Wambura Mwita, Vinai Suriyanon, et al. 2007. ``Time from HIV
Seroconversion to Death: A Collaborative Analysis of Eight Studies in
Six Low and Middle-Income Countries Before Highly Active Antiretroviral
Therapy.'' \emph{AIDS (London, England)} 21 (Suppl 6): S55--S63.
\url{https://doi.org/10.1097/01.aids.0000299411.75269.e8}.

\leavevmode\hypertarget{ref-yiannoutsos_estimated_2012}{}%
Yiannoutsos, Constantin Theodore, Leigh Francis Johnson, Andrew Boulle,
Beverly Sue Musick, Thomas Gsponer, Eric Balestre, Matthew Law, Bryan E
Shepherd, and Matthias Egger. 2012. ``Estimated Mortality of Adult
HIV-Infected Patients Starting Treatment with Combination Antiretroviral
Therapy.'' \emph{Sexually Transmitted Infections} 88 (Suppl\_2):
i33--i43. \url{https://doi.org/10.1136/sextrans-2012-050658}.

\end{document}